\documentclass[preprint,aps]{revtex4}
\usepackage{soul}
\usepackage{graphicx}
\usepackage{epstopdf}
\usepackage{bm}
\usepackage{upgreek}
\usepackage{amsmath}
\usepackage{lipsum}
\usepackage{color,xcolor}
\begin{document}

\title{van Hove Singularity-Driven Emergence of Multiple Flat Bands in Kagome Superconductors}

\author{Hailan Luo$^{1,2\sharp}$, Lin Zhao$^{1,2,3\sharp}$, Zhen Zhao$^{1,2\sharp}$, Haitao Yang$^{1,2,3,4\sharp}$, Yun-Peng Huang$^{1}$, Hongxiong Liu$^{1,2}$, Yuhao Gu$^{1}$, Feng Jin$^{1}$,  Hao Chen$^{1,2}$, Taimin Miao$^{1,2}$, Chaohui Yin$^{1,2}$,   Chengmin Shen$^{1}$, Xiaolin Ren$^{1,2}$, Bo Liang$^{1,2}$, Yingjie Shu$^{1,2}$, Yiwen Chen$^{1,2}$, Fengfeng Zhang$^{5}$, Feng Yang$^{5}$, Shenjin Zhang$^{5}$, Qinjun Peng$^{5}$, Hanqing Mao$^{1}$, Guodong Liu$^{1,2}$, Jiangping Hu$^{1,2}$, Youguo Shi$^{1}$, Zuyan Xu$^{5}$, Kun Jiang$^{1}$, Qingming Zhang$^{1,2}$, Ziqiang Wang$^{6*}$, Hongjun Gao$^{1,2,3,4*}$ and X. J. Zhou$^{1,2,3*}$
}

\affiliation{
\\$^{1}$Beijing National Laboratory for Condensed Matter Physics, Institute of Physics, Chinese Academy of Sciences, Beijing 100190, China
\\$^{2}$University of Chinese Academy of Sciences, Beijing 100049, China
\\$^{3}$Songshan Lake Materials Laboratory, Dongguan 523808, China
\\$^{4}$CAS Center for Excellence in Topological Quantum Computation, University of Chinese Academy of Sciences, Beijing 100190, China.
\\$^{5}$Technical Institute of Physics and Chemistry, Chinese Academy of Sciences, Beijing 100190, China
\\$^{6}$Department of Physics, Boston College, Chestnut Hill, MA 02467, USA
\\$^{\sharp}$These people contributed equally to the present work.
\\$^{*}$Corresponding authors: xjzhou@iphy.ac.cn, hjgao@iphy.ac.cn and wangzi@bc.edu
}

\date{March 09, 2024}


\maketitle

{\bf The newly discovered Kagome superconductors AV$_3$Sb$_5$ (A=K, Rb and Cs) continue to bring surprises in generating unusual phenomena and physical properties, including anomalous Hall effect, unconventional charge density wave, electronic nematicity and time-reversal symmetry breaking. Here we report an unexpected emergence of multiple flat bands in the AV$_3$Sb$_5$ superconductors. By performing high-resolution angle-resolved photoemission (ARPES) measurements, we observed four branches of flat bands that span over the entire momentum space. The appearance of the flat bands is not anticipated from the band structure calculations and cannot be accounted for by the known mechanisms of flat band generation. It is intimately related to the evolution of van Hove singularities. It is for the first time to observe such emergence of multiple flat bands in solid materials. Our findings provide new insights in revealing the underlying mechanism that governs the unusual behaviors in the Kagome superconductors. They also provide a new pathway in producing flat bands and set a platform to study the flat bands related physics.
}

The Kagome lattice, with corner-sharing triangle networks, engenders characteristic electronic structures with van Hove singularities (vHs) at the Brillouin zone boundary, Dirac cone at the zone corner and a flat band through the whole momentum space\cite{MFranz20090520HMGuo,AOBrien20100219PFulde}. Such unique electronic structures facilitate the exploration for a plethora of novel phenomena such as quantum spin liquid phase, non-trivial topology, superconductivity and other correlated phenomena\cite{XGWen20090415WHKo,QHWang20120904WSWang, SRWhite20101130SMYan, MRNorman20161202, TKNg20170418KKanoda,MFranz20090520HMGuo,20160810_HYang2017BHYan,AOBrien20100219PFulde,20110109_GAFirte,20060222_YBKim2006SVIsakov,MFranz20090520HMGuo,20120904_QHWang2013WSWang,20090415_XGWen2009WHKo,20120627_RThomale2012MKiesel,20120904_QHWang2013WSWang,20120919_RThomale2013MLKiesel}. The newly discovered Kagome superconductors AV$_3$Sb$_5$ (A=K, Rb and Cs)\cite{ESToberer20190227BROrtiz,SDWilson20200802BROrtiz} have immediately attracted tremendous interests because they generate abundant quantum states and physical properties including the giant anomalous Hall effect without long-range magnetic order\cite{202007_MNAli2020SYYang,XHChen20210222FHYu}, unconventional charge and pair density waves\cite{MZHasan20201231YXJiang,20210317_HMiao2021HXLi,ZLiang20210308XHChen,HJGao20210929HChen}, electronic nematicity\cite{XHChen20210902LNie}, time-reversal symmetry breaking\cite{ZGuguchia20210625CMielke,MJGraf20210722LYu} and possible unconventional superconductivity\cite{SYLee20210216CCZhao,ZLiang20210308XHChen, HQYuan20210322WYDuan,JLLuo20210609CMu,DLFeng20210417HSXu, ZGuguchia20210625CMielke}. Understanding of these novel quantum phenomena requires thorough investigations on the electronic structures of AV$_3$Sb$_5$ systems. 

On the other hand, the flat band systems have emerged as a fertile playground in condensed matter physics exemplified by the twisted bilayer graphene\cite{PJHerrero20180405YCao,YSWu20140403ZLiu, SFlach20180129DLeykam}. The flat bands are characterized by quenched kinetic energy, singular density of states, electron correlation and localization. Plenty of Coulomb interaction-driven many-body states are expected in flat band systems, such as ferromagnetism\cite{AMielke19901119,HTasaki19920519}, superconductivity\cite{SMiyahara20070410,MPacheco20100424ESMorell}, Wigner crystal\cite{SDSarma20070208CWu} and quantum Hall states\cite{XGWen20110610ETang,CMudry20101222TNeupert,SDSarma20101228KSun,SCZhang20150714GXu}. Theoretically, the flat bands are predicted mainly based on two origins: the localized atomic orbitals and some peculiar lattice geometries\cite{KUeda1997HTsunetsugu,BSutherland19860120, EHLieb19890306,AMielke19901119,HTasaki19920519, MPacheco20100424ESMorell,NRegnault2022BBernevig}. However, the experimental observation of flat bands becomes possible only recently and its realization is limited to a few systems such as heavy fermion materials\cite{MShi20130616NXu}, Kagome compounds\cite{RComin20190605MGKang,JYang2023XJZhou} and twisted bilayer graphene\cite{PJHerrero20180405YCao,MUtama20191028FWang, SLisiF20200212FBaumberger}.

\vspace{3mm}

We carried out angle-resolved photoemission (ARPES) measurements by using our lab-based laser ARPES system (see Methods). We systematically measured AV$_3$Sb$_5$ (A=K, Rb and Cs) (abbreviated as KVS, RVS and CVS, respectively) and Ti-substituted CsV$_{3-x}$Ti$_x$Sb$_5$\,(x=0.04, 0.15 and 0.27) (abbreviated as CVS(0.04), CVS(0.15) and CVS(0.27), respectively) samples (see Methods). AV$_3$Sb$_5$ exhibit charge density wave (CDW) transitions at $\sim$80\,K in KV$_3$Sb$_5$, $\sim$102\,K in RbV$_3$Sb$_5$ and $\sim$93\,K in CsV$_3$Sb$_5$\cite{ESToberer20190227BROrtiz,SDWilson20200802BROrtiz,HCLei20210126QWYin,XJZhou20210714HLLuo,HJGao20211021HTYang}. The Ti-substitution in CsV$_{3-x}$Ti$_x$Sb$_5$ introduces holes and suppresses the CDW transition\cite{HJGao20211021HTYang}. The CsV$_{3-x}$Ti$_x$Sb$_5$\,(x=0.04) exhibits a CDW transition at $\sim$63\,K while CDW transition is undetectable in the CsV$_{3-x}$Ti$_x$Sb$_5$\,(x=0.15 and x=0.27) samples\cite{HJGao20211021HTYang}.

Figure 1 shows band structure of CsV$_3$Sb$_5$ measured along the $\bar{\Gamma}$-$\rm\bar{M}$ high-symmetry direction at 20\,K. In addition to the original $\alpha$ band around $\bar{\Gamma}$ and $\beta$, $\gamma$ and $\delta$ bands around $\rm\bar{M}$\cite{SDWilson20200802BROrtiz,XJZhou20210714HLLuo}, four dispersionless features are clearly observed that span the entire measured momentum space, as marked by arrows and labelled as FB1-FB4 in Fig. 1a. These features show up more clearly in the second-derivative images in Fig. 1b. Their energy positions are at the binding energies of $\sim$70\,meV (FB1), $\sim$200\,meV (FB2), $\sim$550\,meV (FB3) and $\sim$700\,meV (FB4). The presence of these flat bands can also be observed in the photoemission spectra (energy distribution curves, EDCs) (Fig. 1c) and gets more evident in the corresponding second-derivative EDCs (Fig. 1d). 

The observation of these multiple flat bands are unexpected because no such bands are present in the calculated band structures\,\cite{ESToberer20190227BROrtiz, XJZhou20210714HLLuo}. In  our ARPES measurements with $\sim$\,15\,$\mu$m spot size, we observed various regions with different intensities of the folded bands (Fig. S1 in Supplementary Materials) related to the CDW order and surface reconstructions of different cleaved surfaces \cite{XJZhou20210714HLLuo, ZLiang20210308XHChen,JYu2022WLi}. However, the flat bands can be observed both in regions with strong band foldings and in regions with little band foldings (Fig. S1 in Supplementary Materials), indicating that they are not induced by the bulk CDW or surface reconstructions. These flat bands are robust and can be observed in measurements under different polarization geometries (Fig. S2 in Supplementary Materials). Furthermore, they are observed not only along $\bar{\Gamma}$-$\rm\bar{M}$ direction, but also along $\bar{\Gamma}$-$\rm\bar{K}$ and other directions (Fig. S3 in Supplementary Materials), thus covering the entire two-dimensional Brillouin zone.
\vspace{3mm}

The flat bands are ubiquitous in the AV$_3$Sb$_5$ (A=K, Rb and Cs) compounds and show a peculiar doping dependence in CsV$_{3-x}$Ti$_x$Sb$_5$ (x=0$\sim$0.27). Fig. 2 shows the band structures measured at 20\,K along the $\rm\bar{M}$-$\bar{\Gamma}$-$\rm\bar{M}$ direction in KV$_3$Sb$_5$ (Fig. 2a), RbV$_3$Sb$_5$ (Fig. 2b), CsV$_3$Sb$_5$ (Fig. 2c), CsV$_{3-x}$Ti$_x$Sb$_5$(x=0.04) (Fig. 2d), CsV$_{3-x}$Ti$_x$Sb$_5$(x=0.15) (Fig. 2e) and CsV$_{3-x}$Ti$_x$Sb$_5$(x=0.27) (Fig. 2f) samples. The corresponding second-derivative images are shown in Fig. 2g-2l. The EDCs for different samples at a particular momentum away from the main bands are presented in Fig. 2m, with contributions mainly from the flat bands. The corresponding second-derivative EDCs are presented in Fig. 2n. The multiple flat bands are clearly observed in all the parent AV$_3$Sb$_5$ (A=K, Rb and Cs) compounds with similar energy scales. Upon hole doping through Ti-substitution in CsV$_{3-x}$Ti$_x$Sb$_5$, the FB1 and FB2 bands remain pronounced with only a slight energy position variation due to hole doping-induced chemical potential shift. In contrast, the FB3 and FB4 bands exhibit a dramatic change with the hole doping. With increasing hole doping, the FB3 and FB4 bands get weaker in the spectral intensity, shrink in their energy separation and eventually merge into a single flat band FB34 at the high doping in the CVS(0.27) sample. The quantitative energy position variation of the flat bands in these samples is summarized in Fig. 2o. It has been found that the hole doping in CsV$_{3-x}$Ti$_x$Sb$_5$ suppresses the CDW transition which is completely undetectable in the high doping CVS(0.27) sample\cite{HJGao20211021HTYang}. The observed doping evolution of the FB3 and FB4 flat bands is closely related to the doping evolution of the CDW transition in CsV$_{3-x}$Ti$_x$Sb$_5$. 
\vspace{3mm}

In order to further check on the relation between the flat bands and the CDW transition, we carried out temperature-dependent measurements on RbV$_3$Sb$_5$ across the CDW transition temperature of $\sim$102\,K. Fig. 3 shows the band structures of RbV$_3$Sb$_5$ measured along the $\rm\bar{M}$-$\bar{\Gamma}$-$\rm\bar{M}$ direction at different temperatures. The corresponding EDCs and the second-derivative EDCs at a particular momentum are presented in Fig. 3b and 3c, respectively. As seen in Fig. 3, the FB1 and FB2 flat bands are observed in the entire temperature range of 20$\sim$170\,K. Their energy positions show little temperature dependence when crossing the CDW transition. On the other hand, the FB3 and FB4 flat bands exhibit a dramatic evolution with temperature. With the increasing temperature, the spectral intensity of these two flat bands gets weaker, the energy separation between them gets smaller, and eventually these two flat bands FB3 and FB4 merge into one flat band FB34 at temperatures above the CDW transition temperature of $\sim$102\,K. The energy position evolution with temperature for the flat bands is summarized in Fig. 3d. The temperature dependence of the flat bands shows a strong resemblance to the doping dependence shown in Fig. 2. These strongly indicate that the origin of the FB3 and FB4 flat bands is closely related to the CDW transition.

In order to understand the origin of the flat bands, we compared their energy positions with those original bands and found that the emergence of the flat bands is closely related to the van Hove singularities at the $\rm\bar{M}$ point. Fig. 4a shows the band structure of CsV$_3$Sb$_5$ measured along $\bar{\Gamma}$-$\rm\bar{M}$ direction at 20\,K in the CDW state. This measurement covers both the flat bands and the original bands at the $\rm\bar{M}$ point, making their direct comparison possible. As seen from the band structure calculations in Fig. 4b, in the CDW state, three van Hove singularities (vHs1, vHs2 and vHs3) are present at $\rm\bar{M}$. Each van Hove singularity splits in energy and forms an upper branch (vHs1u, vHs2u and vHs3u) and a lower branch (vHs1l, vHs2l and vHs3l). The observed bands at $\rm\bar{M}$ in Fig. 4a show a good correspondence to those in Fig. 4b, as labelled on the right of Fig. 4a. Fig. 4c shows the second-derivative EDCs measured at k$_{\mathrel{/\mskip-2.5mu/}}$\,=\,-0.27 1/$\rm\mathring{A}$ in Fig. 4a highlighting the contribution of the flat bands and at k$_{\mathrel{/\mskip-2.5mu/}}$\,=\,+0.66 1/$\rm\mathring{A}$ in Fig. 4a emphasizing the contribution of the original bands at $\rm\bar{M}$. The energy positions of the flat bands exhibit a good correspondence to those of the van Hove singularity bands. When it comes to the CsV$_{3-x}$Ti$_x$Sb$_5$(x=0.27) sample with high hole-doping without CDW transition, there remains three van Hove singularities expected at $\rm\bar{M}$ (Fig. 4e). However, the vHs3 does not split although vHs1 and vHs2 keep splitting at $\rm\bar{M}$. The expected three bands can be clearly observed in Fig. 4d at $\rm\bar{M}$ as marked on its right side. Fig. 4f shows the second-derivative EDCs measured at k$_{\mathrel{/\mskip-2.5mu/}}$\,=\,-0.27 1/$\rm\mathring{A}$ and at $\rm\bar{M}$ from Fig. 4d. Again, the energy positions of the flat bands shows an excellent agreement with those of the van Hove singularity bands. These results strongly indicate that the emergence of the flat bands is intimately linked to the band structures at $\rm\bar{M}$, particularly the van Hove singularities.

Flat bands systems, constituted by localized or dressed heavy electrons with large effective masses, are an ideal platform to explore for correlated electronic phenomena, such as magnetism and unconventional superconductivity\,\cite{YSWu20140403ZLiu,SWirth2016FSteglich, PJHerrero20180405YCao}. There are several mechanisms that can produce the dispersionless extension of spectral intensity in the momentum space. When electrons are localized within atoms and the electron hopping is limited, a flat band can be formed, such as the 4\emph{f} electrons in heavy Fermion materials (Fig. 4g)\,\cite{SWirth2016FSteglich}. In some special lattice geometries, like in Kagome lattice (Fig. 4h), the flat band can be formed due to destructive interference\,\cite{YSWu20140403ZLiu}. In addition, we note that there are two other ways to produce flat band-like features. In the case of electron coupling with a bosonic mode, a flat band-like feature may appear at the mode energy $\Omega_0$ (Fig. 4i, see Fig. S4 in Supplementary Materials)\,\cite{HYan2023XJZhou}. In the case of electron-mode coupling with the presence of a vHs, a flat band-like feature may emerge at $\Omega_0$+$\emph{E}$ with $\emph{E}$ being the binding energy of the vHs (Fig. 4j, see Fig. S5 in Supplementary Materials)\,\cite{ASandvik2004NBickers}.

Our measured results provide important information on understanding the origin of the flat bands. Since multiple flat bands are observed in AV$_3$Sb$_5$ and their energies coincide with those of vHs, this rules out the possibility that these flat bands come from impurity bands\,\cite{LMiao2018LWray,TYilmaz2020BSinkovic,APertsova2021ABalatsky}. This can also exclude the possibility of localized atomic orbital bands (Fig. 4g). One possibility is whether these flat bands may come from the scattering of electrons at the vHs by impurities or disorders. To the best of our knowledge, such phenomena have not been observed in the previous ARPES measurements of materials with vHs \cite{KGofron1994BDabrowski,RComin20190605MGKang}. Although the cleaved sample surface consists of different regions, like Cs-terminated, half Cs-terminated and Sb-terminated in CsV$_3$Sb$_5$\cite {ZLiang20210308XHChen}, we always observe similar flat bands, even in regions with strong band foldings and in regions with little band foldings (Fig. S1 in Supplementary Materials). These results indicate that the flat bands we observed are more likely attributed to the intrinsic bulk properties of AV$_3$Sb$_5$ materials.

Our observed flat bands are different from the Kagome lattice-induced ones (Fig. 4h) which are $\sim$1\,eV away from the Fermi level\,\cite{RComin2021MGKang}. Electron-mode coupling can produce flat band-like feature (Fig. 4i). However, our observation of four flat bands is unlikely due to such electron-mode coupling because no collective excitations with similar energy scales have been observed in AV$_3$Sb$_5$. Furthermore, in the electron-mode coupling picture (Fig. 4i), the spectral weight of the flat band-like feature decays rapidly with the momentum moving away from the main band. This is inconsistent with our observation that the spectral weight of the flat bands changes little in the wide momentum space. In the case of electron-mode coupling with a coexisting vHs, the flat band-like feature can also occur near the energy position of the vHs (Fig. 4j). This scenario closely connects the flat band-like feature with the vHs which is consistent with our observations. But similar to the electron-mode coupling case, the flat band-like feature loses its spectral weight rapidly with the momentum moving away from the vHs. This is not consistent with the nearly uniform momentum distribution of the flat bands we observed.

In summary, we have discovered the emergence of multiple flat bands in AV$_3$Sb$_5$. These observations are unexpected and the origin of these flat bands can not be understood by the known mechanisms of flat band generation. The occurrence of these flat bands is clearly associated with the van Hove singularities which may be further related to their singular density of states (DOS). The van Hove singularities have been reported in a number of systems, but the vHs-driven generation of the flat bands has been observed for the first time in AV$_3$Sb$_5$. These indicate that some unusual mechanisms are at play in AV$_3$Sb$_5$, like the discovery of the unexpected anomalous Hall effect in this system. Our work provides a new paradigm to generate flat bands and will stimulate further efforts to understand its origin and other unusual phenomena in Kagome superconductors.
\vspace{3mm}

\noindent{\bf Methods}

\noindent{\bf Growth and characterization of single crystals.} High quality single crystals of AV$_3$Sb$_5$ (A=K, Rb and Cs) and CsV$_{3-x}$Ti$_x$Sb$_5$ (x=0.04, 0.15 and 0.27) were grown from modified self-flux methods\,\cite{ESToberer20190227BROrtiz,HJGao20211021HTYang}. The crystals were characterized by X-ray diffraction and measurements of the magnetic susceptibilty and electrical resistance\,\cite{XJZhou20210714HLLuo,HJGao20211021HTYang}. AV$_3$Sb$_5$ exhibit CDW transitions at $\sim$80\,K (KV$_3$Sb$_5$), $\sim$102\,K (RbV$_3$Sb$_5$) and $\sim$93\,K (CsV$_3$Sb$_5$)\,\cite{ESToberer20190227BROrtiz, SDWilson20200802BROrtiz, HCLei20210126QWYin, XJZhou20210714HLLuo, HJGao20211021HTYang}, respectively. The Ti-substititon for V in CsV$_{3-x}$Ti$_x$Sb$_5$ crystal has been confirmed by the scanning transmission electron microscopy, electron energy-loss spectroscopy and scanning tunneling microscope\,\cite{HJGao20211021HTYang}. CsV$_{3-x}$Ti$_x$Sb$_5$\,(x=0.04) exhibits a CDW transition at $\sim$63\,K while no CDW transition is observed in the CsV$_{3-x}$Ti$_x$Sb$_5$\,(x=0.15 and x=0.27) samples\,\cite{HJGao20211021HTYang}.
\vspace{3mm}

\noindent{\bf High resolution ARPES measurements} High-resolution angle-resolved photoemission measurements were carried out on our lab system equipped with a Scienta DA30L electron energy analyzer\,\cite{20071016_XJZhou2008GDLiu,XJZhou2018}. We use ultraviolet laser as the light source that can provide a photon energy of \emph{h}$\nu$=6.994\,eV with a bandwidth of 0.26\,meV. The energy resolution was set at $\sim$2.5\,meV. The angular resolution is $\sim$0.3 $^\circ$, corresponding to a momentum resolution of $\sim$0.004\,1/$\rm\mathring{A}$ for the photon energy of 6.994\,eV. The laser spot size is set at $\sim$15\,$\rm\mu$m. The Fermi level is referenced by measuring on a clean polycrystalline gold that is electrically connected to the sample. The sample was cleaved \emph{in situ} and measured in vacuum with a base pressure better than 5$\times$10$^{-11}$\,Torr.
\vspace{3mm}

\noindent{\bf Band structure calculations.} First-principles calculations are performed by using the Projected Augmented Wave Method (PAW) density functional theory (DFT), as implemented in the Vienna Ab Initio Simulation Package (VASP)\,\cite{19650621_LJSham1965WKohn,19940613_PEBlochl1994,19961213_JFurthmuller1996GKresse}. A 2$\times$2$\times$1 supercell is constructed to describe the TrH CDW phase of AV$_3$Sb$_5$. The crystal structures are relaxed by using the Perdew-Burke-Ernzerhof (PBE) functional\,\cite{19960521_MErnzerhof1996JPPerdew} and zero damping DFT-D3 van der Waals correction\,\cite{20100118_HKrieg2010SGrimme} until the forces are less than 0.001 eV/$\rm\mathring{A}$. The cutoff energy of the plane wave basis is set at 600\,eV and the energy convergence criterion is set at 10$^{-7}$\,eV. The corresponding Brillouin zones are sampled by using a 16$\times$16$\times$10 (for primitive cell) and a 8$\times$8$\times$10 (for supercell) ${\Gamma}$ centered $\bf{k}$-grid. The effective band structure is calculated by the band-unfolding method\,\cite{20100416_AZunger2010VPopescu,20111021_AZunger2012VPopescu} proposed by Zunger et al. with BandUP code\,\cite{2014_Medeiros,2015_Medeiros}. The 2$\times$2$\times$1 TrH CDW order reconstruction is an input in our band-unfolding calculation. All the DFT calculations are \textit{ab-initio} without any adjustable parameters except for the standard exchange-correlation functional and pseudopotentials.
\vspace{3mm}

\noindent{\bf Spectral function simulation of the electron-mode coupling.} We performed a simulation of the single-particle spectral function considering the electron coupling with a bosonic mode. The purpose here is to illustrate the emergence of the flat band-like feature in the case of electron-mode coupling with a coexisting vHs. We follow the fashion of the Migdal theory\,\cite{AMigdal1957} and calculate the Migdal electron-mode self-energy $\Sigma_{ep}$. 

\begin{equation}
    \Sigma_{ep}\left(\mathbf{k},i\omega_n\right)=\frac{1}{N}\sum_{\mathbf{q}}|g_{\mathbf{k},\mathbf{q}}|^2
    \left(
      \frac{b\left(\Omega_\mathbf{q}\right)+f\left(\epsilon_{\mathbf{k+q}}\right)}{i\omega_n+\Omega_\mathbf{q}-\epsilon_{\mathbf{k+q}}}
    + \frac{1+b\left(\Omega_\mathbf{q}\right)-f\left(\epsilon_{\mathbf{k+q}}\right)}{i\omega_n-\Omega_\mathbf{q}-\epsilon_{\mathbf{k+q}}}
    \right),
\end{equation}
For simplicity, we considered a case of Einstein mode with constant frequency $\Omega_\mathbf{q}\equiv\Omega_0$ and homogeneous electron-mode coupling constant $g_{\mathbf{k},\mathbf{q}}\equiv g$. Here, $b\left(\Omega\right)$ is the Bose-Einstein distribution function, and $f\left(\epsilon_{\mathbf{k}}\right)$ is the Fermi-Dirac distribution. The spectral function is given by $A\left(\mathbf{k},\omega\right)=-\frac{1}{\pi}\mathrm{Im} \, G\left(\mathbf{k},i\omega_n\rightarrow\omega+i\delta\right)$, where the Green function $G\left(\mathbf{k},i\omega_n\right)=\left(i\omega_n-\epsilon_\mathbf{k}-\Sigma_{ep}\right)^{-1}$. In the simulation, we take the case of square lattice, and the dispersion $\epsilon_{\mathbf{k}}=2t\left(\cos k_x+\cos k_y\right)$. The cut is along the high symmetry line $\Gamma-\mathrm{X}-\mathrm{M}$. The vHs at $\mathrm{X}=\left(\pi,0\right)$ is placed at the energy of $-0.4\,\mathrm{eV}$. The mode frequency $\Omega_0$ is set to be $30\,\mathrm{meV}$ and the coupling constant is $g=0.2$. The simulated results are shown in Fig. S5 in Supplementary Materials and in Fig. 4j.
\vspace{3mm}

\noindent {\bf Acknowledgement}\\
This work is supported by the National Natural Science Foundation of China (Grant Nos. 11888101, 61888102, 11974404, 12074411 and U2032204), the National Key Research and Development Program of China (Grant Nos. 2021YFA1401800, 2022YFA1403900 and 2018YFA0704200), the Strategic Priority Research Program (B) of the Chinese Academy of Sciences (Grant Nos. XDB25000000, XDB28000000 and XDB33000000), the Innovation Program for
Quantum Science and Technology (Grant No. 2021ZD0301800), the Youth Innovation Promotion Association of CAS (Grant No. Y2021006), the U.S. Department of Energy,
Basic Energy Sciences (Grant No. DE-FG02-99ER45747), the K. C. Wong Education Foundation (GJTD-2018-01), and the Synergetic Extreme Condition User Facility (SECUF).

\noindent {\bf Author Contributions}\\
X.J.Z., H.J.G., L.Z. and H.L.L. conceived this project. H.L.L. performed ARPES experiments and analyzed the ARPES data.  Z.Z., H.T.Y., C.M.S. and H.J.G. contributed to crystal growth of pure CsV$_3$Sb$_5$ and Ti-doped CsV$_3$Sb$_5$. H.X.L. and Y.G.S. contributed to crystal growth of KV$_3$Sb$_5$, RbV$_3$Sb$_5$ and pure CsV$_3$Sb$_5$. Y.P.H., Y.H.G., J.P.H., K.J. and Z.Q.W. contributed to DFT calculations and theoretical analysis. F.J. and Q.M.Z. contributed to Raman measurements and analysis. L.Z., H.C., C.H.Y., T.M.M., X.L.R., B.L., Y.J.S., Y.W.C., H.Q.M., G.D.L. and X.J.Z. contributed to the development and maintenance of the ARPES systems and related software development. H.L.L., F.F.Z., F.Y., S.J.Z., Q.J.P. and Z.Y.X. contributed to the maintenance of 7\,eV laser system. H.L.L., L.Z. and X.J.Z. wrote this paper with information from all co-authors. All authors participated in discussion and comment on the paper.

\newpage

\begin{figure*}[tbp]
\begin{center}
\includegraphics[width=1\columnwidth,angle=0]{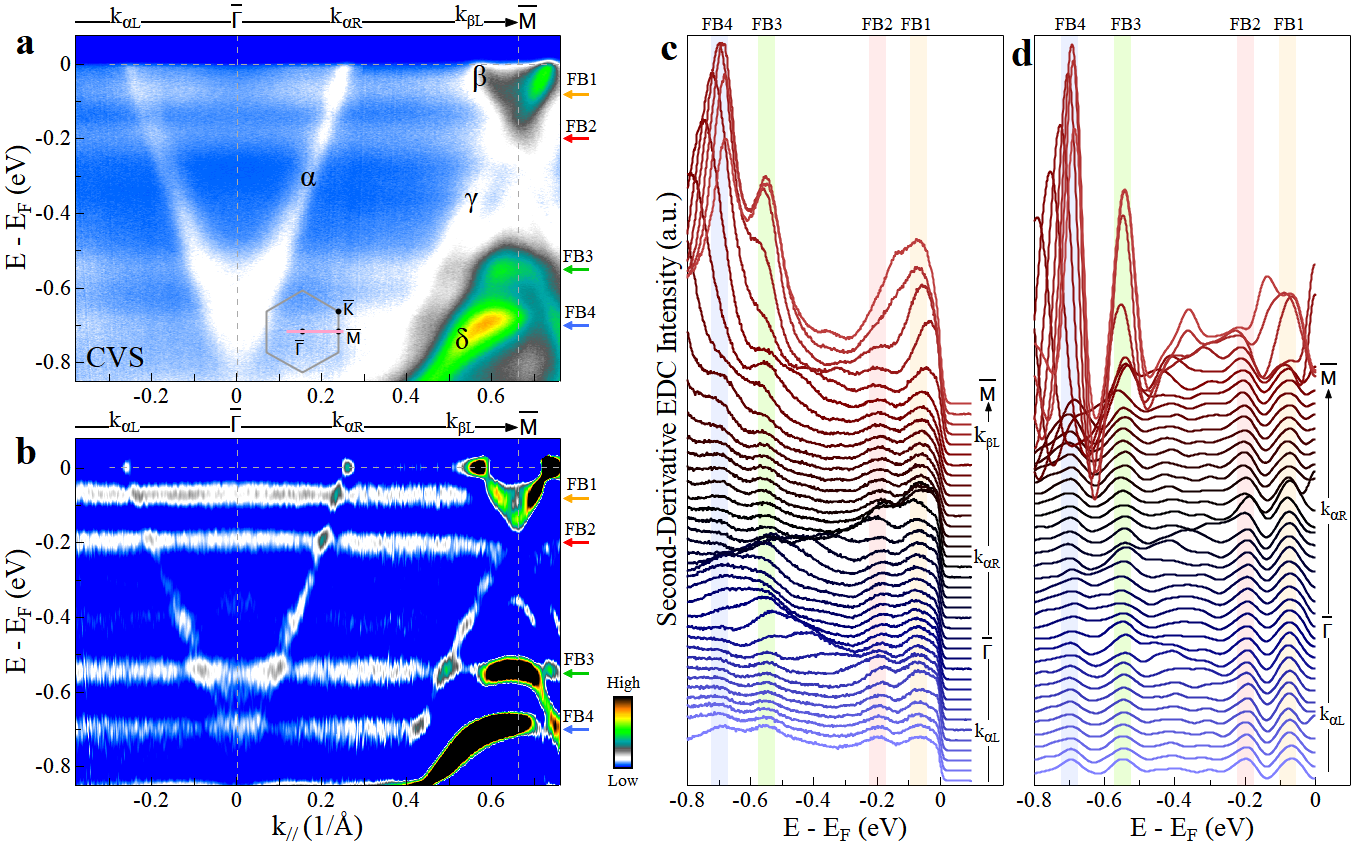}
\end{center}
\caption{\textbf{Observation of multiple flat bands over the entire momentum space in CsV$_3$Sb$_5$.} {\textbf{a,}} Band structure measured at 20\,K along the $\bar{\Gamma}$-$\rm\bar{M}$ direction. The location of the momentum cut is marked as a solid pink line in the inset. {\textbf{b,}} The corresponding second-derivative image with respect to energy obtained from (a). The four arrows on the right side of (a) and (b) are guides to the eye for the observed flat bands. {\textbf{c,}} The corresponding photoemission spectra (energy distribution curves, EDCs) from the band structure in (a). The momentum range where the EDCs are extracted is marked as a black arrow on the top of (a). The EDCs extracted at the Fermi momenta (k$_{\alpha}${$\rm_L$}, k$_{\alpha}${$\rm_R$} and k$_{\beta}${$\rm_L$}) and high-symmetry points ($\bar{\Gamma}$ and $\rm\bar{M}$) are marked. Four vertical shaded regions at the binding energies of 70\,meV (FB1), 200\,meV (FB2), 550\,meV (FB3) and 700\,meV (FB4) are guides to the eye for the flat bands. {\textbf{d,}} The corresponding second derivative EDCs from the second derivative image in (b). The momentum range where the second derivative EDCs are extracted is marked as a black arrow on the top of (b).
}
\end{figure*}

\begin{figure*}[tbp]
\begin{center}
\includegraphics[width=1\columnwidth,angle=0]{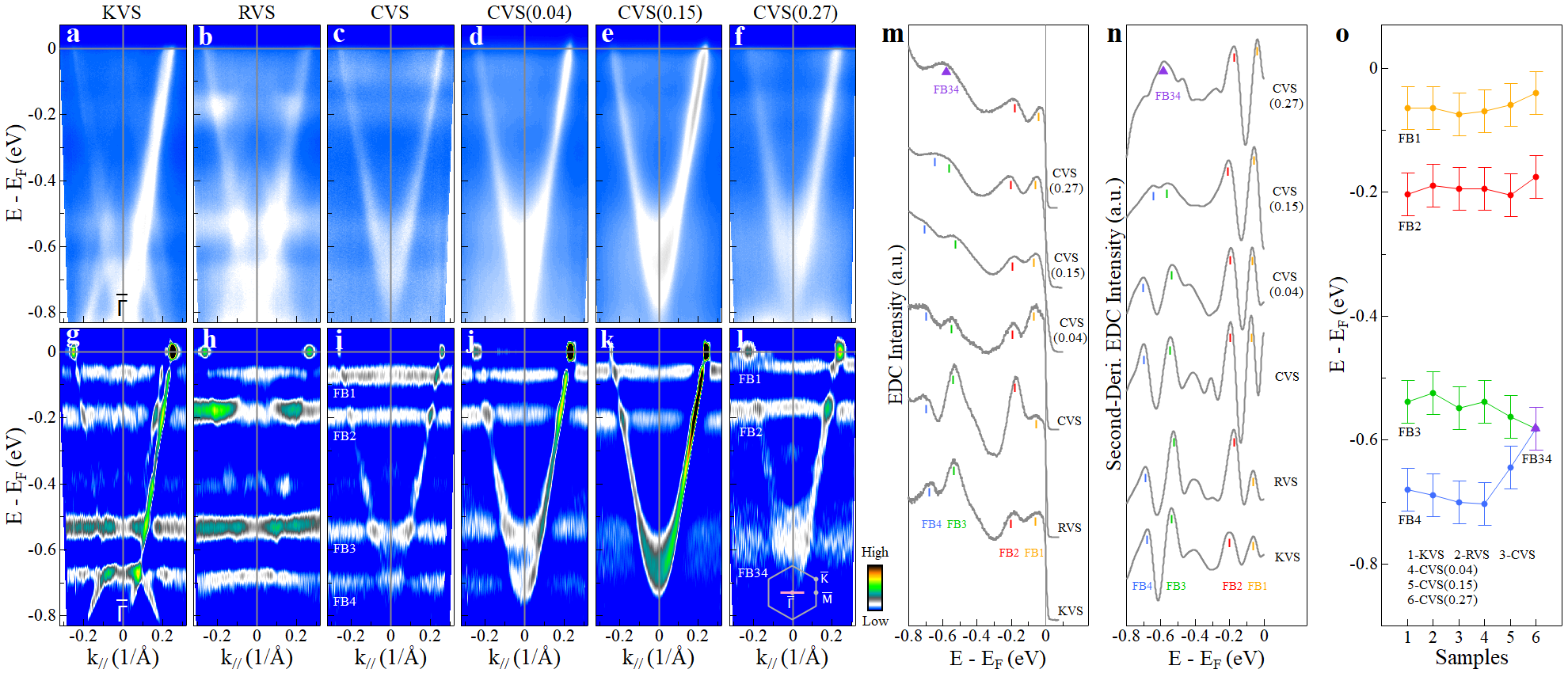}
\end{center}
\caption{\textbf{Ubiquitous flat bands in AV$_3$Sb$_5$ (A\,=\,K, Rb and Cs) and their doping dependence.} {\textbf{a-f,}} Band structures measured at 20\,K along the $\rm\bar{M}$-$\bar{\Gamma}$-$\rm\bar{M}$ direction in KV$_3$Sb$_5$ (KVS) (a), RbV$_3$Sb$_5$ (RVS) (b), CsV$_3$Sb$_5$ (CVS) (c), CsV$_{3-x}$Ti$_x$Sb$_5$(x=0.04) (CVS(0.04)) (d), CsV$_{3-x}$Ti$_x$Sb$_5$(x=0.15) (CVS(0.15)) (e) and  CsV$_{3-x}$Ti$_x$Sb$_5$(x=0.27) (CVS(0.27)) (f). The location of the momentum cuts for (a)-(f) is marked as a solid pink line in the inset of (l). {\textbf{g-l,}} The corresponding second-derivative images with respect to energy obtained from (a)-(f). {\textbf{m,}} EDCs extracted at k$_{\mathrel{/\mskip-2.5mu/}}$\,=\,-0.27 1/$\rm\AA$ from (a-f). This momentum point is chosen to highlight the contribution from the flat bands because it is away from the original and folded bands. FB1-FB4 and FB34 mark the position of the flat bands. In each EDC, the position of the flat bands are marked by bars or triangles. {\textbf{n,}} Second derivative EDCs extracted at k$_{\mathrel{/\mskip-2.5mu/}}$\,=\,-0.27 1/$\rm\AA$ from (g-l). {\textbf{o,}} Sample-dependent and doping-dependent energy positions of the observed flat bands. Samples 1-6 represent KVS, RVS, CVS, CVS(0.04), CVS(0.15) and CVS(0.27), respectively, as marked on the bottom. The error bars reflect the uncertainty in determining the energy positions of the flat bands.
}
\end{figure*}

\begin{figure*}[tbp]
\begin{center}
\includegraphics[width=1\columnwidth,angle=0]{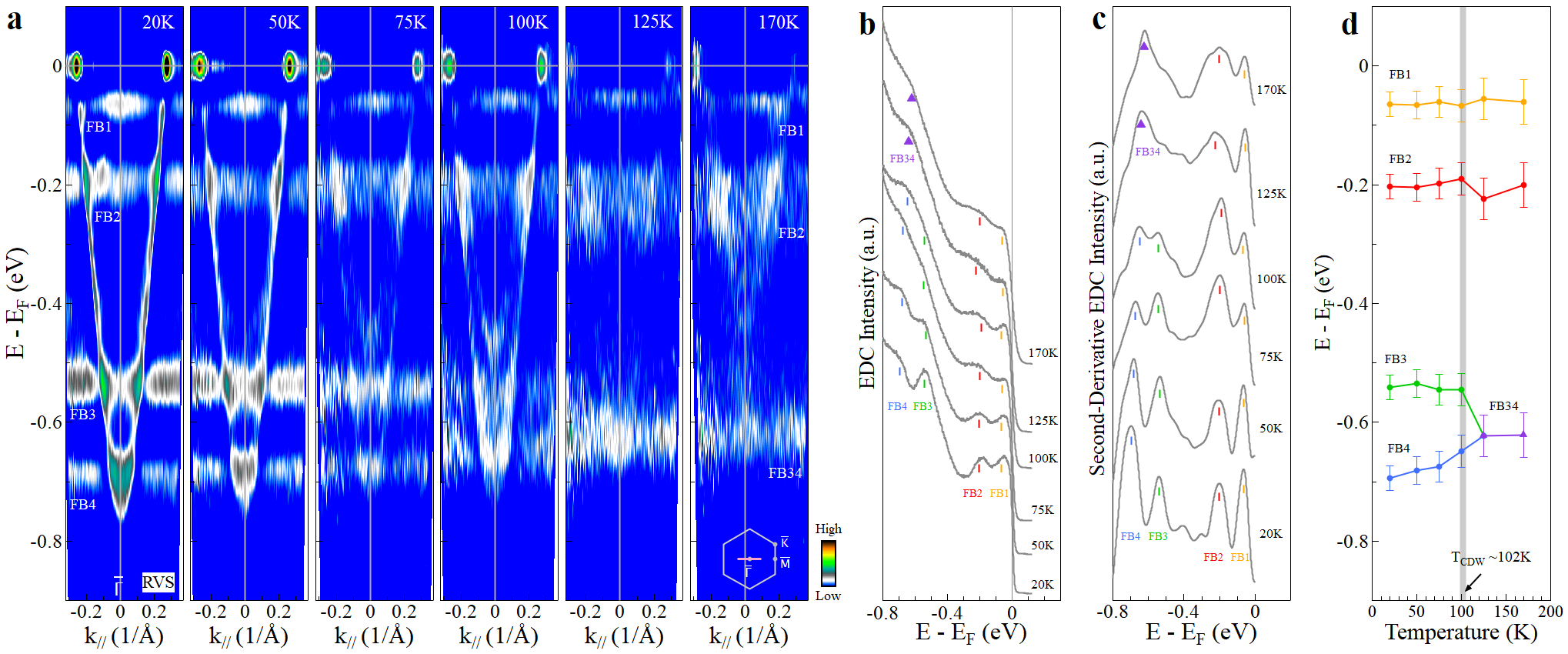}
\end{center}
\caption{\textbf{Evolution of the flat bands with temperature in RbV$_3$Sb$_5$.} {\textbf{a,}} Band structures measured at different temperatures from 20\,K to 170\,K along the $\rm\bar{M}$-$\bar{\Gamma}$-$\rm\bar{M}$ direction. The location of the momentum cut is marked as a solid pink line in the inset of the right-most panel. These are second derivative images with respect to energy obtained from the original data. {\textbf{b,}} The EDCs at k$_{\mathrel{/\mskip-2.5mu/}}$\,=\,+0.30 1/$\rm\AA$ measured at different temperatures extracted from the original data. {\textbf{c,}} The corresponding second-derivative EDCs at k$_{\mathrel{/\mskip-2.5mu/}}$\,=\,+0.30 1/$\rm\AA$ measured at different temperatures extracted from (a). FB1-FB4 and FB34 indicate the energy positions of the flat bands. In each EDC, the positions of the flat bands are marked by bars or triangles in different colors. {\textbf{d,}} Temperature-dependent energy positions of the flat bands. The error bars reflect the uncertainty in determining the energy positions of the flat bands.
}
\end{figure*}

\begin{figure*}[tbp]
\begin{center}
\includegraphics[width=0.8\columnwidth,angle=0]{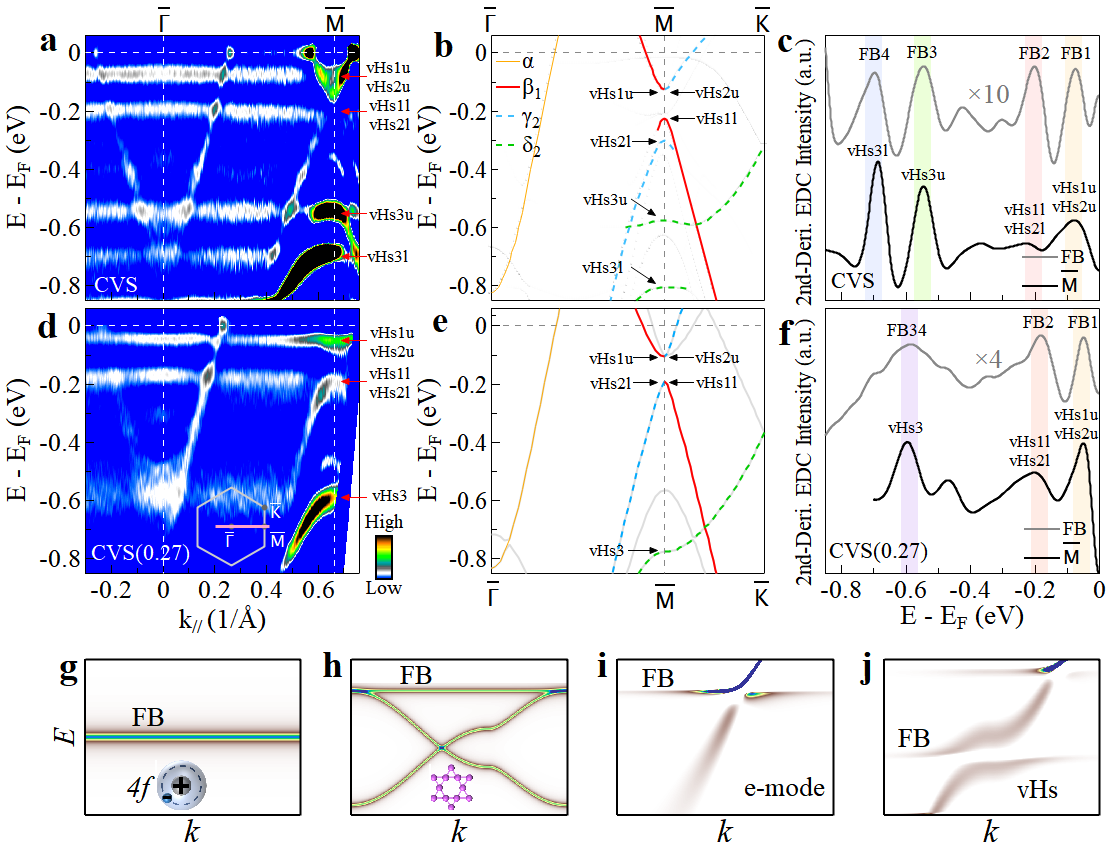}
\end{center}
\caption{\textbf{The relation between the flat bands and the van Hove singularities and the origins of flat bands.} {\textbf{a and d,}} Band structures measured at 20\,K along the $\bar{\Gamma}$-$\rm\bar{M}$ direction in CsV$_3$Sb$_5$ (a) and CsV$_{3-x}$Ti$_x$Sb$_5$ (x=0.27) (d). The location of the momentum cuts is marked as a solid  pink line in the inset of (d). These are second-derivative images with respect to energy obtained from original data. {\textbf{b and e,}} Calculated band structures along the $\bar{\Gamma}$-$\rm\bar{M}$-$\rm\bar{K}$ direction in CsV$_3$Sb$_5$ with the reconstructed TrH crystal structure\cite{20210415_SDWilson2021BROrtiz} (b) and with pristine crystal structure (e) at  k$\rm_z$\,=\,$\pi$/c without spin-orbit coupling. The van Hove singularities (vHs1, vHs2 and vHs3) are marked by arrows which are associated with the $\beta${$_1$}, $\gamma${$_2$} and $\delta${$_2$} bands, respectively, that are highlighted by colored lines. When a van Hove singularity splits at $\rm\bar{M}$, the two branches are labelled with suffixes u (up) and d (down) such as vHs3u and vHs3d. {\textbf{c,}} Second-derivative EDCs at k$_{\mathrel{/\mskip-2.5mu/}}$\,=\,+0.30 1/$\rm\mathring{A}$(grey line) contributed mainly from the flat bands and at $\rm\bar{M}$ (black line) contributed mainly from the van Hove singularities, extracted from (a). For comparison, the second-derivative EDC at k$_{\mathrel{/\mskip-2.5mu/}}$\,=\,+0.30 1/$\rm\mathring{A}$ is multiplied by 10 in intensity. {\textbf{f,}} Same as (c) but extracted from (d). \textbf{g,} Flat bands in heavy fermion materials. \textbf{h,} The flat band in Kagome lattice. \textbf{i,} The flat band-like feature in electron-mode coupling\,\cite{HYan2023XJZhou}. This is the EDC second derivative image; the original image is shown in Fig. S4 in Supplementary Materials. \textbf{j,} The flat band-like feature coming from van Hove singularity in an electron-mode coupling system. This is the EDC second derivative image; the original image is shown in Fig. S5 in Supplementary Materials.
}
\end{figure*}


\begin{thebibliography}{10}

\bibitem{MFranz20090520HMGuo}
H.~M. Guo and M.~Franz.
\newblock {Topological insulator on the kagome lattice}.
\newblock {\em Physical Review B}, 80(11):113102, 2009.

\bibitem{AOBrien20100219PFulde}
A.~O’Brien, F.~Pollmann, and P.~Fulde.
\newblock {Strongly correlated fermions on a kagome lattice}.
\newblock {\em Physical Review B}, 81(23):235115, 2010.

\bibitem{XGWen20090415WHKo}
W.~H. Ko, P.~A. Lee, and X.~G. Wen.
\newblock {Doped kagome system as exotic superconductor}.
\newblock {\em Physical Review B}, 79(21):214502, 2009.

\bibitem{QHWang20120904WSWang}
W.~S. Wang, Z.~Z. Li, Y.~Y. Xiang, and Q.~H. Wang.
\newblock {Competing electronic orders on kagome lattices at van Hove filling}.
\newblock {\em Physical Review B}, 87(11):115135, 2013.

\bibitem{SRWhite20101130SMYan}
S.~Yan, D.~A. Huse, and S.~R. White.
\newblock {Spin-liquid ground state of the S = 1/2 kagome Heisenberg
  antiferromagnet}.
\newblock {\em Science}, 332:1173--1176, 2011.

\bibitem{MRNorman20161202}
M. R. Norman.
\newblock {Colloquium: Herbertsmithite and the search for the quantum spin liquid}.
\newblock {\em Reviews of Modern Physics}, 88(4):041002, 2016.

\bibitem{TKNg20170418KKanoda}
Y.~Zhou, K.~Kanoda, and T.~K. Ng.
\newblock {Quantum spin liquid states}.
\newblock {\em Reviews of Modern Physics}, 89(2):025003, 2017.

\bibitem{20160810_HYang2017BHYan}
H.~Yang, Y.~Sun, Y.~Zhang, W.~J. Shi, S.~S.~P. Parkin, and B.~Yan.
\newblock {Topological Weyl semimetals in the chiral antiferromagnetic
  materials Mn$_3$Ge and Mn$_3$Sn}.
\newblock {\em New Journal of Physics}, 19(1):015008, 2017.

\bibitem{20110109_GAFirte}
A.~R$\rm\ddot{u}$egg and G.~A. Fiete.
\newblock {Fractionally charged topological point defects on the kagome
  lattice}.
\newblock {\em Physical Review B}, 83(16):165118, 2011.

\bibitem{20060222_YBKim2006SVIsakov}
S.~V. Isakov, S.~Wessel, R.~G. Melko, K.~Sengupta, and Y.~B. Kim.
\newblock {Hard-Core bosons on the kagome lattice: valence-bond solids and
  their quantum melting}.
\newblock {\em Physical Review Letters}, 97(14):147202, 2006.

\bibitem{20120904_QHWang2013WSWang}
W.~S. Wang, Z.~Z. Li, Y.~Y. Xiang, and Q.~H. Wang.
\newblock {Competing electronic orders on kagome lattices at van Hove filling}.
\newblock {\em Physical Review B}, 87(11):115135, 2013.

\bibitem{20090415_XGWen2009WHKo}
W.~H. Ko, P.~A. Lee, and X.~G. Wen.
\newblock {Doped kagome system as exotic superconductor}.
\newblock {\em Physical Review B}, 79(21):214502, 2009.

\bibitem{20120627_RThomale2012MKiesel}
M.~L. Kiesel and R.~Thomale.
\newblock {Sublattice interference in the kagome Hubbard model}.
\newblock {\em Physical Review B}, 86(12):121105, 2012.

\bibitem{20120919_RThomale2013MLKiesel}
M.~L. Kiesel, C.~Platt, and R.~Thomale.
\newblock {Unconventional Fermi surface instabilities in the kagome Hubbard
  model}.
\newblock {\em Physical Review Letters}, 110(12):126405, 2013.

\bibitem{ESToberer20190227BROrtiz}
B.~R. Ortiz, L.~C. Gomes, J.~R. Morey, M.~Winiarski, M.~Bordelon, J.~S. Mangum,
  I.~W.~H. Oswald, J.~A. Rodriguez-Rivera, J.~R. Neilson, S.~D. Wilson,
  E.~Ertekin, T.~M. McQueen, and E.~S. Toberer.
\newblock {New kagome prototype materials: discovery of KV$_3$Sb$_5$,
  RbV$_3$Sb$_5$, and CsV$_3$Sb$_5$}.
\newblock {\em Physical Review Materials}, 3(9):094407, 2019.

\bibitem{SDWilson20200802BROrtiz}
B.~R. Ortiz, S.~M.~L. Teicher, Y.~Hu, J.~L. Zuo, P.~M. Sarte, E.~C. Schueller,
  A.~M.~M. Abeykoon, M.~J. Krogstad, S.~Rosenkranz, R.~Osborn, R.~Seshadri,
  L.~Balents, J.~He, and S.~D. Wilson.
\newblock {CsV$_3$Sb$_5$: a Z$_2$ topological kagome metal with a
  superconducting ground state}.
\newblock {\em Physical Review Letters}, 125(24):247002, 2020.

\bibitem{202007_MNAli2020SYYang}
S.~Yang, Y.~Wang, B.~R. Ortiz, D.~Liu, J.~Gayles, E.~Derunova,
  R.~Gonzalez-Hernandez, L.~$\rm\check{S}$mejkal, Y.~Chen, S.~S.~P. Parkin,
  S.~D. Wilson, E.~S. Toberer, T.~McQueen, and M.~N. Ali.
\newblock {Giant, unconventional anomalous Hall effect in the metallic
  frustrated magnet candidate, KV$_3$Sb$_5$}.
\newblock {\em Science Advances}, 6(31):eabb6003, 2020.

\bibitem{XHChen20210222FHYu}
F.~H. Yu, T.~Wu, Z.~Y. Wang, B.~Lei, W.~Z. Zhuo, J.~J. Ying, and X.~H. Chen.
\newblock {Concurrence of anomalous Hall effect and charge density wave in a
  superconducting topological kagome metal}.
\newblock {\em Physical Review B}, 104(4):L041103, 2021.

\bibitem{MZHasan20201231YXJiang}
Y.~X. Jiang, J.~X. Yin, M.~M. Denner, N.~Shumiya, B.~R. Ortiz, G.~Xu,
  Z.~Guguchia, J.~He, M.~S. Hossain, X.~Liu, J.~Ruff, L.~Kautzsch, S.~S. Zhang,
  G.~Chang, I.~Belopolski, Q.~Zhang, T.~A. Cochran, D.~Multer, M.~Litskevich,
  Z.~J. Cheng, X.~P. Yang, Z.~Wang, R.~Thomale, T.~Neupert, S.~D. Wilson, and
  M.~Z. Hasan.
\newblock {Unconventional chiral charge order in kagome superconductor
  KV$_3$Sb$_5$}.
\newblock {\em Nature Materials}, 20:1353--1357, 2021.

\bibitem{20210317_HMiao2021HXLi}
H. Li, T. T. Zhang, T. Yilmaz, Y. Y. Pai, C. E. Marvinney, A. Said, Q. W. Yin, C. S. Gong, Z. J. Tu, E. Vescovo, C. S. Nelson, R. G. Moore, S. Murakami, H. C. Lei, H. N. Lee, B. J. Lawrie, and H. Miao.

\newblock {Observation of unconventional charge density wave without acoustic
  phonon anomaly in kagome superconductors ${A\mathrm{V}}_{3}{\mathrm{Sb}}_{5}$
  ($A=\mathrm{Rb}$, Cs)}.
\newblock {\em Physical Review X}, 11(3):031050, 2021.

\bibitem{ZLiang20210308XHChen}
Z.~Liang, X.~Hou, F.~Zhang, W.~Ma, P.~Wu, Z.~Zhang, F.~Yu, J.~J. Ying,
  K.~Jiang, L.~Shan, Z.~Wang, and X.~H. Chen.
\newblock {Three-dimensional charge density wave and surface-dependent
  vortex-core states in a kagome superconductor
  ${\mathrm{CsV}}_{3}{\mathrm{Sb}}_{5}$}.
\newblock {\em Physical Review X}, 11(3):031026, 2021.

\bibitem{HJGao20210929HChen}
H.~Chen, H.~Yang, B.~Hu, Z.~Zhao, J.~Yuan, Y.~Xing, G.~Qian, Z.~Huang, G.~Li,
  Y.~Ye, S.~Ma, S.~Ni, H.~Zhang, Q.~Yin, C.~Gong, Z.~Tu, H.~Lei, H.~Tan,
  S.~Zhou, C.~Shen, X.~Dong, B.~Yan, Z.~Wang, and H.~J. Gao.
\newblock {Roton pair density wave in a strong-coupling kagome superconductor}.
\newblock {\em Nature}, 599:222--228, 2021.

\bibitem{XHChen20210902LNie}
L.~Nie, K.~Sun, W.~Ma, D.~Song, L.~Zheng, Z.~Liang, P.~Wu, F.~Yu, J.~Li,
  M.~Shan, D.~Zhao, S.~Li, B.~Kang, Z.~Wu, Y.~Zhou, K.~Liu, Z.~Xiang, J.~Ying,
  Z.~Wang, T.~Wu, and X.~Chen.
\newblock {Charge-density-wave-driven electronic nematicity in a kagome
  superconductor}.
\newblock {\em Nature}, 604:59--64, 2022.

\bibitem{ZGuguchia20210625CMielke}
C.~Mielke, D.~Das, J.~X. Yin, H.~Liu, R.~Gupta, Y.~X. Jiang, M.~Medarde, X.~Wu,
  H.~C. Lei, J.~Chang, Pengcheng Dai, Q.~Si, H.~Miao, R.~Thomale, T.~Neupert,
  Y.~Shi, R.~Khasanov, M.~Z. Hasan, H.~Luetkens, and Z.~Guguchia.
\newblock {Time-reversal symmetry-breaking charge order in a kagome
  superconductor}.
\newblock {\em Nature}, 602:245--250, 2022.

\bibitem{MJGraf20210722LYu}
L.~Yu, C.~Wang, Y.~Zhang, M.~Sander, S.~Ni, Z.~Lu, S.~Ma, Z.~Wang, Z.~Zhao,
  H.~Chen, K.~Jiang, Y.~Zhang, H.~Yang, F.~Zhou, X.~Dong, S.~L. Johnson, M.~J.
  Graf, J.~Hu, H.~J. Gao, and Z.~Zhao.
\newblock {Evidence of a hidden flux phase in the topological kagome metal
  CsV$_3$Sb$_5$}.
\newblock arXiv:2107.10714, 2021.

\bibitem{SYLee20210216CCZhao}
C.~C. Zhao, L.~S. Wang, W.~Xia, Q.~W. Yin, J.~M. Ni, Y.~Y. Huang, C.~P. Tu,
  Z.~C. Tao, Z.~J. Tu, C.~S. Gong, H.~C. Lei, Y.~F. Guo, X.~F. Yang, and S.~Y.
  Li.
\newblock {Nodal superconductivity and superconducting domes in the topological
  kagome metal CsV$_3$Sb$_5$}.
\newblock arXiv:2102.08356, 2021.

\bibitem{HQYuan20210322WYDuan}
W.~Duan, Z.~Nie, S.~Luo, F.~Yu, B.~R. Ortiz, L.~Yin, H.~Su, F.~Du, A.~Wang,
  Y.~Chen, X.~Lu, J.~Ying, S.~D. Wilson, X.~Chen, Y.~Song, and H.~Yuan.
\newblock {Nodeless superconductivity in the kagome metal CsV$_3$Sb$_5$}.
\newblock {\em Science China Physics, Mechanics  Astronomy}, 64(10):107462,
  2021.

\bibitem{JLLuo20210609CMu}
C.~Mu, Q.~Yin, Z.~Tu, C.~Gong, H.~Lei, Z.~Li, and J.~Luo.
\newblock {S-Wave superconductivity in kagome metal CsV$_3$Sb$_5$ revealed by
  $^{121/123}$Sb NQR and $^{51}$V NMR measurements}.
\newblock {\em Chinese Physics Letters}, 38(7):077402, 2021.

\bibitem{DLFeng20210417HSXu}
H.~S. Xu, Y.~J. Yan, R.~Yin, W.~Xia, S.~Fang, Z.~Chen, Y.~Li, W.~Yang, Y.~Guo,
  and D.~L. Feng.
\newblock {Multiband superconductivity with sign-preserving order parameter in
  kagome superconductor ${\mathrm{CsV}}_{3}{\mathrm{Sb}}_{5}$}.
\newblock {\em Physical Review Letters}, 127(18):187004, 2021.

\bibitem{PJHerrero20180405YCao}
Y.~Cao, V.~Fatemi, S.~Fang, K.~Watanabe, T.~Taniguchi, E.~Kaxiras, and
  P.~Jarillo-Herrero.
\newblock {Unconventional superconductivity in magic-angle graphene
  superlattices}.
\newblock {\em Nature}, 556:43--50, 2018.

\bibitem{YSWu20140403ZLiu}
Z.~Liu, F.~Liu, and Y.~S. Wu.
\newblock {Exotic electronic states in the world of flat bands: From theory to
  material}.
\newblock {\em Chinese Physics B}, 23(7):077308, 2014.

\bibitem{SFlach20180129DLeykam}
D.~Leykam, A.~Andreanov, and S.~Flach.
\newblock {Artificial flat band systems: from lattice models to experiments}.
\newblock {\em Advances in Physics: X}, 3(1):1473052, 2018.

\bibitem{AMielke19901119}
A.~Mielke.
\newblock {Ferromagnetic ground states for the Hubbard model on line graphs}.
\newblock {\em Journal of Physics A: Mathematical and General}, 24(2):L73--L77,
  1991.

\bibitem{HTasaki19920519}
H.~Tasaki.
\newblock {Ferromagnetism in the Hubbard models with degenerate single-electron
  ground states}.
\newblock {\em Physical Review Letters}, 69(10):1608--1611, 1992.

\bibitem{SMiyahara20070410}
S.~Miyahara, S.~Kusuta, and N.~Furukawa.
\newblock {BCS theory on a flat band lattice}.
\newblock {\em Physica C: Superconductivity}, 460-462:1145--1146, 2007.

\bibitem{MPacheco20100424ESMorell}
E.~Su$\rm\acute{a}$rez~Morell, J.~D. Correa, P.~Vargas, M.~Pacheco, and
  Z.~Barticevic.
\newblock {Flat bands in slightly twisted bilayer graphene: Tight-binding
  calculations}.
\newblock {\em Physical Review B}, 82(12):121407, 2010.

\bibitem{SDSarma20070208CWu}
C.~Wu, D.~Bergman, L.~Balents, and D.~S. Sarma.
\newblock {Flat bands and Wigner crystallization in the honeycomb optical
  lattice}.
\newblock {\em Physical Review Letters}, 99(7):070401, 2007.

\bibitem{XGWen20110610ETang}
E.~Tang, J.~W. Mei, and X.~G. Wen.
\newblock {High-temperature fractional quantum hall states}.
\newblock {\em Physical Review Letters}, 106(23):236802, 2011.

\bibitem{CMudry20101222TNeupert}
T.~Neupert, L.~Santos, C.~Chamon, and C.~Mudry.
\newblock {Fractional quantum hall states at zero magnetic field}.
\newblock {\em Physical Review Letters}, 106(23):236804, 2011.

\bibitem{SDSarma20101228KSun}
K.~Sun, Z.~Gu, H.~Katsura, and S.~D Sarma.
\newblock {Nearly flatbands with nontrivial topology}.
\newblock {\em Physical Review Letters}, 106(23):236803, 2011.

\bibitem{SCZhang20150714GXu}
G.~Xu, B.~Lian, and S.~C. Zhang.
\newblock {Intrinsic quantum anomalous hall effect in the kagome lattice
  ${\mathrm{Cs}}_{2}{\mathrm{LiMn}}_{3}{\mathrm{F}}_{12}$}.
\newblock {\em Physical Review Letters}, 115(18):186802, 2015.

\bibitem{KUeda1997HTsunetsugu}
H.~Tsunetsugu, M.~Sigrist, and K.~Ueda.
\newblock {The ground-state phase diagram of the one-dimensional Kondo lattice
  model}.
\newblock {\em Reviews of Modern Physics}, 69(3):809--864, 1997.

\bibitem{BSutherland19860120}
B.~Sutherland.
\newblock {Localization of electronic wave functions due to local topology}.
\newblock {\em Physical Review B}, 34(8):5208--5211, 1986.

\bibitem{EHLieb19890306}
E.~H. Lieb.
\newblock {Two theorems on the Hubbard model}.
\newblock {\em Physical Review Letters}, 62(10):1201--1204, 1989.

\bibitem{NRegnault2022BBernevig}
N.~Regnault, Y.~Xu, M.~R. Li, D.~S. Ma, M.~Jovanovic, A.~Yazdani, S.~S. Parkin,
  C.~Felser, L.~M. Schoop, N.~P. Ong, R.~J. Cava, L.~Elcore, Z.~D. Song, and
  B.~A. Bernevig.
\newblock {Catalogue of flat-band stoichiometric materials}.
\newblock {\em Nature}, 603:824--828, 2022.

\bibitem{MShi20130616NXu}
N.~Xu, X.~Shi, P.~K. Biswas, C.~E. Matt, R.~S. Dhaka, Y.~Huang, N.~C. Plumb,
  M.~Radović, J.~H. Dil, E.~Pomjakushina, K.~Conder, A.~Amato, Z.~Salman,
  D.~McK Paul, J.~Mesot, H.~Ding, and M.~Shi.
\newblock {Surface and bulk electronic structure of the strongly correlated
  system SmB${}_{6}$ and implications for a topological Kondo insulator}.
\newblock {\em Physical Review B}, 88(12):121102, 2013.

\bibitem{RComin20190605MGKang}
M.~Kang, L.~Ye, S.~Fang, J.~S. You, A.~Levitan, M.~Han, J.~I. Facio,
  C.~Jozwiak, A.~Bostwick, E.~Rotenberg, M.~K. Chan, R.~D. McDonald, D.~Graf,
  K.~Kaznatcheev, E.~Vescovo, D.~C. Bell, E.~Kaxiras, J.~van~den Brink,
  M.~Richter, M.~Prasad~Ghimire, J.~G. Checkelsky, and R.~Comin.
\newblock {Dirac fermions and flat bands in the ideal kagome metal FeSn}.
\newblock {\em Nature Materials}, 19(2):163--169, 2020.

\bibitem{JYang2023XJZhou}
J.~Yang, X.~Yi, Z.~Zhao, Y.~Xie, T.~Miao, H.~Luo, H.~Chen, B.~Liang, W.~Zhu,
  Y.~Ye, J.~Y. You, B.~Gu, S.~Zhang, F.~Zhang, F.~Yang, Z.~Wang, Q.~Peng,
  H.~Mao, G.~Liu, Z.~Xu, H.~Chen, H.~Yang, G.~Su, H.~Gao, and X.~J. Zhou.
\newblock {Observation of flat band, Dirac nodal lines and topological surface
  states in Kagome superconductor CsTi$_3$Bi$_5$}.
\newblock {\em Nature Communication}, 14:4089, 2023.

\bibitem{MUtama20191028FWang}
M.~I.~B. Utama, R.~J. Koch, K.~Lee, N.~Leconte, H.~Li, S.~Zhao, L.~Jiang,
  J.~Zhu, K.~Watanabe, T.~Taniguchi, P.~D. Ashby, A.~Weber-Bargioni, A.~Zettl,
  C.~Jozwiak, J.~Jung, E.~Rotenberg, A.~Bostwick, and F.~Wang.
\newblock {Visualization of the flat electronic band in twisted bilayer
  graphene near the magic angle twist}.
\newblock {\em Nature Physics}, 17(2):184--188, 2021.

\bibitem{SLisiF20200212FBaumberger}
S.~Lisi, X.~Lu, T.~Benschop, T.~A. de~Jong, P.~Stepanov, J.~R. Duran,
  F.~Margot, I.~Cucchi, E.~Cappelli, A.~Hunter, A.~Tamai, V.~Kandyba,
  A.~Giampietri, A.~Barinov, J.~Jobst, V.~Stalman, M.~Leeuwenhoek, K.~Watanabe,
  T.~Taniguchi, L.~Rademaker, S.~J. van~der Molen, M.~P. Allan, D.~K. Efetov,
  and F.~Baumberger.
\newblock {Observation of flat bands in twisted bilayer graphene}.
\newblock {\em Nature Physics}, 17(2):189--193, 2021.

\bibitem{HCLei20210126QWYin}
Q.~Yin, Z.~Tu, C.~Gong, Y.~Fu, S.~Yan, and H.~Lei.
\newblock {Superconductivity and normal-state properties of kagome metal
  RbV$_3$Sb$_5$ single crystals}.
\newblock {\em Chinese Physics Letters}, 38(3):037403, 2021.

\bibitem{XJZhou20210714HLLuo}
H.~Luo, Q.~Gao, H.~Liu, Y.~Gu, D.~Wu, C.~Yi, J.~Jia, S.~Wu, X.~Luo, Y.~Xu,
  L.~Zhao, Q.~Wang, H.~Mao, G.~Liu, Z.~Zhu, Y.~Shi, K.~Jiang, J.~Hu, Z.~Xu, and
  X.~J. Zhou.
\newblock {Electronic nature of charge density wave and electron-phonon
  coupling in kagome superconductor KV$_3$Sb$_5$}.
\newblock {\em Nature Communications}, 13(1):273, 2022.

\bibitem{HJGao20211021HTYang}
H.~Yang, Y.~Zhang, Z.~Huang, Z.~Zhao, J.~Shi, G.~Qian, B.~Hu, Z.~Lu, H.~Zhang,
  C.~Shen, X.~Lin, Z.~Wang, S.~J. Pennycook, H.~Chen, X.~Dong, W.~Zhou, and
  H.~J. Gao.
\newblock {Titanium doped kagome superconductor CsV$_{3-x}$Ti$_x$Sb$_5$ and two distinct phases}.
\newblock {\em Science Bulletin}, 67(21):2176--2185, 2022.

\bibitem{JYu2022WLi}
J.~Yu, Z.~Xu, K.~Xiao, Y.~Yuan, Q.~Yin, Z.~Hu, C.~Gong, Y.~Guo, Z.~Tu, P.~Tang,
  H.~Lei, Q.~K. Xue, , and W.~Li.
\newblock {Evolution of electronic structure in pristine and Rb-reconstructed
  surfaces of kagome metal ${\mathrm{RbV}}_{3}{\mathrm{Sb}}_{5}$}.
\newblock {\em Nano Letters}, 22(3):918--925, 2022.

\bibitem{SWirth2016FSteglich}
S.~Wirth and F.~Steglich.
\newblock {Exploring heavy fermions from macroscopic to microscopic length
  scales}.
\newblock {\em Nature Review Physics}, 1(16051), 2016.

\bibitem{HYan2023XJZhou}
H.~Yan, J.~M. Bok, J.~He, W.~Zhang, Q.~Gao, X.~Luo, Y.~Cai, Y.~Peng, J.~Meng,
  C.~Li, H.~Chen, C.~Song, C.~Yin, T.~Miao, Y.~Chen, G.~Gu, C.~Lin, F.~Zhang,
  F.~Yang, S.~Zhang, Q.~Peng, G.~Liu, H.~Y. Zhao, L.and~Choi, Z.~Xu, and X.~J.
  Zhou.
\newblock {Ubiquitous coexisting electron-mode couplings in high-temperature
  cuprate superconductors}.
\newblock {\em Proceedings of the National Academy of Sciences of the United
  States of America}, 120(43):e2219491120, 2023.

\bibitem{ASandvik2004NBickers}
A.~W. Sandvik, D.~J. Scalapino, and N.~E. Bickers.
\newblock {Effect of an electron-phonon interaction on the one-electron
  spectral weight of a \emph{p}-wave superconductor}.
\newblock {\em Physical Review B}, 69:094523, 2004.

\bibitem{LMiao2018LWray}
L.~Miao, Y.~Xu, W.~Zhang, D.~Older, S.~A. Breitweiser, E.~Kotta, H.~He,
  T.~Suzuki, J.~D. Denlinger, R.~R. Biswas, J.~G. Checkelsky, W.~Wu, and L.~A.
  Wray.
\newblock {Observation of a topological insulator Dirac cone reshaped by
  non-magnetic impurity resonance}.
\newblock {\em npj Quantum Materials}, 3(29), 2018.

\bibitem{TYilmaz2020BSinkovic}
T.~Yilmaz, A.~Pertsova, W.~Hines, E.~Vescovo, K.~Kaznatcheev, A.~V. Balatsky,
  and B.~Sinkovic.
\newblock {Gap-like feature observed in the non-magnetic topological
  insulators}.
\newblock {\em Journal of Physics: Condensed Matter}, 32(145503), 2020.

\bibitem{APertsova2021ABalatsky}
A.~Pertsova, P.~Johnson, D.~P. Arovas, and A.~V. Balatsky.
\newblock {Dirac node engineering and flat bands in doped Dirac materials}.
\newblock {\em Physical Review Research}, 3:033001, 2021.

\bibitem{KGofron1994BDabrowski}
K.~Gofron, J.~C. Campuzano, A.~A. Abrikosov, M.~Lindroos, A.~Bansil, H.~Ding,
  Koelling, and B.~Dabrowski.
\newblock {Observation ofan "Extended" Van Hove Singularity in
  YBa$_2$Cu$_4$O$_8$ by Ultrahslh Energy Resolution Angle-Resolved
  Photoemteion}.
\newblock {\em Physical Review Letters}, 73(24):3302--3305, 1994.

\bibitem{RComin2021MGKang}
M.~Kang, S.~Fang, J.~K. Kim, B.~R. Ortiz, J.~Yoo, B.~G. Park, S.~D. Wilson,
  J.~H. Park, and R.~Comin.
\newblock {Twofold van Hove singularity and origin of charge order in
  topological kagome superconductor CsV$_3$Sb$_5$}.
\newblock {\em Nature Physics}, 18:301--308, 2022.

\bibitem{20071016_XJZhou2008GDLiu}
G.~Liu, G.~Wang, Y.~Zhu, H.~Zhang, G.~Zhang, X.~Wang, Y.~Zhou, W.~Zhang,
  H.~Liu, L.~Zhao, J.~Meng, X.~Dong, C.~Chen, Z.~Xu, and X.~J. Zhou.
\newblock {Development of a vacuum ultraviolet laserbased angle-resolved
  photoemission system with a superhigh energy resolution better than 1 meV}.
\newblock {\em Review of Scientific Instruments}, 79(2):023105, 2008.

\bibitem{XJZhou2018}
X.~J. Zhou, S.~He, G.~Liu, L.~Zhao, L.~Yu, and W.~Zhang.
\newblock {New developments in laser-based photoemission spectroscopy and its
  scientific applications: a key issues review}.
\newblock {\em {Reports on Progress in Physics}}, 81(6):062101, 2018.

\bibitem{19650621_LJSham1965WKohn}
W.~Kohn and L.~J. Sham.
\newblock {Self-consistent equations including exchange and correlation
  effects}.
\newblock {\em Physical Review}, 140(4A):A1133--A1138, 1965.

\bibitem{19940613_PEBlochl1994}
P.~E. Bl{\"o}chl.
\newblock {Projector augmented-wave method}.
\newblock {\em Physical Review B}, 50(24):17953--17979, 1994.

\bibitem{19961213_JFurthmuller1996GKresse}
G.~Kresse and J.~Furthm{\"u}ller.
\newblock {Efficient iterative schemes for ab initio total-energy calculations
  using a plane-wave basis set}.
\newblock {\em Physical Review B}, 54(16):11169--11186, 1996.

\bibitem{19960521_MErnzerhof1996JPPerdew}
J.~P. Perdew, K.~Burke, and M.~Ernzerhof.
\newblock {Generalized gradient approximation made simple}.
\newblock {\em Physical Review Letters}, 77(18):3865--3868, 1996.

\bibitem{20100118_HKrieg2010SGrimme}
S.~Grimme, J.~Antony, S.~Ehrlich, and H.~Krieg.
\newblock {A consistent and accurate ab initio parametrization of density
  functional dispersion correction (DFT-D) for the 94 elements H-Pu}.
\newblock {\em J. Chem. Phys.}, 132(15):154104, 2010.

\bibitem{20100416_AZunger2010VPopescu}
V.~Popescu and A.~Zunger.
\newblock {Effective band structure of random alloys}.
\newblock {\em Physical Review Letters}, 104(23):236403, 2010.

\bibitem{20111021_AZunger2012VPopescu}
V.~Popescu and A.~Zunger.
\newblock {Extracting E versus k effective band structure from supercell
  calculations on alloys and impurities}.
\newblock {\em Physical Review B}, 85(8):085201, 2012.

\bibitem{2014_Medeiros}
P.~V.~C. Medeiros, S.~Stafstr{\"o}m, and J.~Bj{\"o}rk.
\newblock Effects of extrinsic and intrinsic perturbations on the electronic
  structure of graphene: Retaining an effective primitive cell band structure
  by band unfolding.
\newblock {\em Physical Review B}, 89(4):041407, 2014.

\bibitem{2015_Medeiros}
P.~V.~C. Medeiros, S.~S. Tsirkin, S.~Stafstr{\"o}m, and J.~Bj{\"o}rk.
\newblock Unfolding spinor wave functions and expectation values of general
  operators: introducing the unfolding-density operator.
\newblock {\em Physical Review B}, 91(4):041116, 2015.

\bibitem{AMigdal1957}
A.~B. Migdal.
\newblock {Interaction between electrons and lattice vibrations in a normal
  metal}.
\newblock {\em Soviet Physics JETP}, 34(7):6, 1958.

\bibitem{20210415_SDWilson2021BROrtiz}
B.~R. Ortiz, S.~M.~L. Teicher, L.~Kautzsch, P.~M. Sarte, J.~P.~C. Ruff,
  R.~Seshadri, and S.~D. Wilson.
\newblock {Fermi surface mapping and the nature of charge density wave order in
  the kagome superconductor CsV$_3$Sb$_5$}.
\newblock {\em Physical Review X}, 11:041030, 2021.

\end{thebibliography}
\end{document}